\documentclass[fleqn,twoside]{article}
\usepackage{espcrc2,epsfig}

\usepackage{graphicx}
\usepackage[figuresright]{rotating}

\newcommand{\AmS}{{\protect\the\textfont2
  A\kern-.1667em\lower.5ex\hbox{M}\kern-.125emS}}
\newcommand{\figwidth}{6.85cm}

\title{The determination of $\alpha_s$ from lattice QCD with 2+1 flavors of dynamical quarks}

\author{C. Davies\address[UG]{Department of Physics and Astronomy, 
        University of Glasgow, Glasgow G12 8QQ, UK},
        A. Gray\addressmark[UG],
        M. Alford\addressmark[UG],
	E. Follana\addressmark[UG],
        J. Hein\address[EU]{School of Physics, University of Edinburgh, Edinburgh EH9 3JZ, UK},
	P. Lepage\address[CU]{Newman Laboratory of Nuclear Studies, Cornell University, Ithaca NY 14853, USA},
	Q. Mason\addressmark[CU],
	M.Nobes\address[SFU]{Department of Physics, Burnaby, British Columbia V5A 1S6, CANADA},
	J.Shigemitsu\address[OSU]{Department of Physics, the Ohio State University, OH 43210, USA},
	H. Trottier\addressmark[SFU],
	M. Wingate\addressmark[OSU], HPQCD and UKQCD collaborations.}
       
\begin{document}

\begin{abstract}
We describe the first lattice determination of the strong coupling constant 
with 3 flavors of dynamical quarks. The method follows previous analyses in 
using a perturbative expansion for the plaquette and Upsilon spectroscopy to set 
the scale. Using dynamical configurations from the MILC collaboration with 
2+1 flavors of dynamical quarks we are able to avoid previous problems of haivng 
to extrapolate to 3 light flavors from 0 and 2. Our result agrees with our 
previous work: $\alpha^{\overline{MS}}_s(M_Z) = 0.121(3)$. 
\vspace{1pc}
\end{abstract}

\maketitle

\section{INTRODUCTION}

Precise determinations of the strong coupling constant are important for 
strong interaction phenomenology and searches for new physics. Lattice 
QCD provides a good methodology for such a determination~\cite{ouralpha}. 

\section{IMPROVED STAGGERED QUARKS}

The staggered quark formulation allows for much faster simulation of 
dynamical quarks than other formulations. It maintains a remnant chiral symmetry 
at the cost of not entirely removing the doublers, which appear as 
additional `tastes' of quarks. The presence of the doublers
gives rise to large lattice artefacts through taste-changing interactions. 
The realisation that these can be systematically reduced by a Symanzik improvement 
programme opens the way to generating ensembles with light dynamical quarks and 
small lattice artefacts. The MILC collaboration have generated a number 
of dynamical ensembles using the Asqtad action for the dynamical quarks and 
either 2 dynamical flavors, 3 degenerate dynamical flavors or 
2 degenerate plus 1 heavier dynamical flavors, which should be close to 
the real world~\cite{MILC}. In the 2+1 flavor case they have been able to reduce the 
light dynamical mass to 1/5 of the heavier one (which is approximately at $m_s$). 

The Asqtad staggered action includes a set of staples in place of the 
simple 1-link derivative, to suppress gluon exchange with momentum 
$\pi/a$ which couples the tastes together. These staples include a correction 
for low-momentum $a^2$ artefacts introduced by this procedure.  
In addition there is a 3-link derivative (Naik) term to 
improve rotational discretisation errors through $a^2$. The Asqtad action 
then has $\alpha_sa^2$ rotational errors and $\alpha_s^2a^2$ taste-symmetry 
breaking interactions. This is combined in MILC ensembles with a gluon 
action improved to the level of $\alpha_s^2a^2$ discretisation errors. 

We have calculated the $\Upsilon$ spectrum on these ensembles using the 
NRQCD action for the $b$ quark~\cite{agray}. 

\section{NRQCD}

The physics of the bottomonium ($\Upsilon$) system  
is that of non-relativistic $b$ quarks ($v^2/c^2 \approx$ 10\%) and 
so the NRQCD action 
is appropriate~\cite{NRQCD}. We use a Hamiltonian correct through 
$a^2v^4/c^4$ where all terms are tadpole-improved using an estimate of the 
Landau gauge link~\cite{agray}. For the purposes of determining the 
lattice spacing to set the scale for $\alpha_s$ we need only the 
radial and orbital excitation energies. These are experimentally well-determined
and have the advantage that they are to a good approximation independent of 
heavy quark mass so that sophisticated tuning of the $b$ mass is not 
necessary. In addition they are good probes of the dynamical quark 
content of the theory because the $\Upsilon$ has no valence 
light quarks. 

The leading contribution to 
the  radial and orbital splittings is at $\cal{O}$$(v^2/c^2)$ and so the expected systematic 
error from higher order (e.g. $\alpha_sv^4/c^4$ or $v^6/c^6$) terms 
not included in the NRQCD action is expected to be at the 1\% level. 
The splittings can be determined very precisely on the lattice because 
of the speed of NRQCD correlator calculations.  

\section{DETERMINATION OF $\alpha_s$}

We use the perturbative expansion of the plaquette in terms of a 
physically well-motivated coupling, $\alpha_V(q)$, defined from the heavy 
quark potential: 
\begin{equation}
V(q) = - \frac{16\pi}{3q^2} \alpha_V(q) .
\end{equation}
We then define $\alpha_P(q)$ by the following expansion of the 
logarithm of the plaquette, where $\alpha_P$ is equal to $\alpha_V$ 
to the order (second) to which we are working here. 
\begin{eqnarray}
-{\rm ln}(plaq) = 3.0682(2)\alpha_P(q^*)[1+ \\\nonumber
\alpha_P\{-0.770(4)-0.09681(9)n_f\}]
\end{eqnarray}
The numbers in brackets indicate numerical uncertainties in the 
perturbative coefficients from the numerical integration involved in their 
calculation. 
This perturbative expansion is for 1-loop Symanzik-improved gluon action 
using tadpole-improvement with $u_0$ from the plaquette and $n_f$ 
flavors of Aqtad improved staggered quarks, as appropriate to the MILC 
ensembles. The scale, $q^*$, is set using the BLM scheme and is 
$3.33/a$ in this case. The perturbative expansion is well-behaved and shows 
small corrections for non-zero $n_f$. 

$\alpha_P$ can be converted to $\alpha_{\overline{MS}}$ using the 
equality of $\alpha_P$ and $\alpha_V$ and the continuum 
relation between $\alpha_V$ and $\alpha_{\overline{MS}}$. 
\begin{equation}
\alpha_{\overline{MS}}^{(n_f)}(q) = \alpha_P^{(n_f)}(e^{5/6}q)[1+2\frac{\alpha_P}{\pi}+{\cal O}(\alpha_P^2)]
\end{equation}

\begin{figure}
\centerline{\epsfig{file=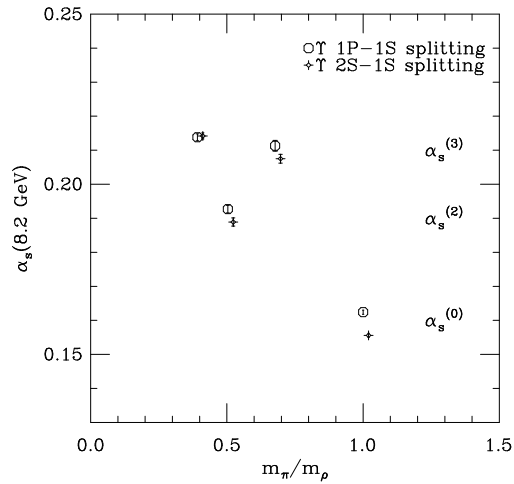,width=\figwidth}}
\caption{$\alpha_P$(8.2GeV) for various MILC ensembles with 
differing numbers of dynamical flavors. Two different quantities 
are used to set the scale: the $\Upsilon$ 1P-1S splitting and 
the $\Upsilon$ 2S-1S splitting. Errors from setting the scale 
are included.
}
\label{fig:alphaunq}
\end{figure}

\section{RESULTS}

Figure 1 shows results for $\alpha_P$ evolved to a common 
(and arbitrarily chosen) physical scale 
of 8.2 GeV for various MILC ensembles. 
Notice the clear separation of results for $n_f = 0$, 2 and 3. 

The figure shows results using either the $\Upsilon$ 
orbital splitting or radial splitting to set the lattice 
spacing. The orbital splitting is calculated from the 
energy difference between the $1^1P_1$ ($h_b$) and $1^3S_1$ ($\Upsilon$)
states on the lattice. The experimental result for 
the $1^1P_1$ mass is not known and so is taken as the spin-average of 
the masses of the $1^3P$ ($\chi_b$) states. The 
difference 
is expected to be very small.  
The radial splitting is the difference in energies of 
the $2^3S_1$ ($\Upsilon'$) and $1^3S_1$ states, which 
is experimentally well-known. 

The value for $\alpha_P(8.2GeV)$ agrees well 
between the two determinations on the 2+1 dynamical ensemble 
but not on the others. This reflects the fact that the 
lattice spacing will depend on the quantity chosen to fix it 
if the vacuum does not have dynamical content which is close enough 
to the `real world'. 

We can take the result for $\alpha_P$ on the 2+1 dynamical 
ensemble, convert to $\alpha_{\overline{MS}}$ and run to 
$M_Z$. The result for $\alpha_{\overline{MS}}^{(5)}$ is 
then 0.121(3). The dominant source of error (0.002) is the perturbative 
error 
from unknown higher orders in perturbation theory in eq. 3.  
This result agrees with the world average and with our previous 
results using dynamical (unimproved) staggered quarks~\cite{ouralpha}. 

Other sources of systematic error give small effects. 
We can see from Figure 1 that the $n_f$ = 3 results show very 
little dependence on dynamical light quark mass. Extrapolating 
$\alpha_P$ with $m_{\pi}^2$ would change the final result by 
less than 0.001. 

The lattice spacing dependence of the result is expected to 
be very small but we currently have $\Upsilon$ results 
only on the coarser of the MILC set of ensembles. 
We can estimate the dependence from other results, however. 
Figure 2 shows the dependence of $\alpha_P$ in the quenched 
approximation on $a^2$ for improved and unimproved glue~\cite{steveg}. 
Here we have used $r_0$ to set the scale (and quote $\alpha_P$ 
at a scale of $15/r_0$) because it can be fixed very precisely. 
Its disadvantage is that it is not clear how to convert it 
to a physical result. The figure shows that unimproved glue has 
very small dependence on $a$ and improved glue (appropriate 
to the MILC ensembles) even less. 
We fold this lattice 
spacing dependence along with quark mass effects above and NRQCD systematics into a 
final error of 0.003. 

\begin{figure}
\centerline{\epsfig{file=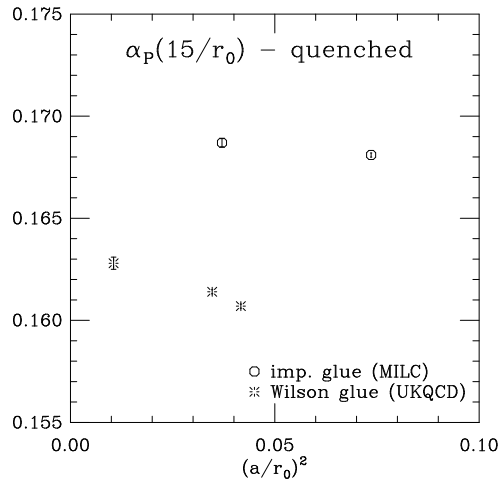,width=\figwidth}}
\caption{$\alpha_P$ in the quenched approximation at a scale of 
$15/r_0$ for improved and unimproved glue.}
\label{fig:alphaq}
\end{figure}

\section{CONCLUSIONS}

The first calculation of $\alpha_s$ with 2+1 flavours of dynamical 
quarks gives results in agreement with previous work, but without 
extrapolating in $n_f$. Further 
improvements, including the calculation of the next order in 
perturbation theory, are planned to reduce the error even 
further. 

\vspace{2mm}
\begin{flushleft}
{\bf Acknowledgements} 

We are grateful to the MILC collaboration for the use of their configurations. 
This work was supported by the DoE, the EU IHP programme, 
NSF, NSERC and PPARC. 

\end{flushleft}


\begin{thebibliography}{9}
\bibitem{ouralpha} C. Davies {\it et al}, Phys. Rev. D{\bf 56} (1997) 2755. 
\bibitem{MILC} MILC collaboration, Phys. Rev. D{\bf 64} (2001) 054506. 
\bibitem{agray} A. Gray {\it et al}, these Proceedings.  
\bibitem{NRQCD} G. P. Lepage {\it et al}, Phys. Rev. D{\bf 46} (1992) 4052. 
\bibitem{steveg} We thank Steve Gottlieb for providing the 
unpublished MILC results for this plot. 
\end{thebibliography}
\end{document}